\documentclass[runningheads,a4paper]{llncs}

\usepackage{amssymb}
\usepackage{graphicx}
\usepackage{float}
\usepackage{amsmath}
\usepackage{xcolor}

\usepackage{url}
\newcommand{\keywords}[1]{\par\addvspace\baselineskip
\noindent\keywordname\enspace\ignorespaces#1}

\begin{document}

\mainmatter  % start of an individual contribution

% first the title is needed
\title{Deep Learning Under Siege: Identifying Security Vulnerabilities and Risk Mitigation Strategies}

% a short form should be given in case it is too long for the running head
\titlerunning{Security Challenges in Deep Learning Architectures}

% the name(s) of the author(s) follow(s) next
%
% NB: Chinese authors should write their first names(s) in front of
% their surnames. This ensures that the names appear correctly in
% the running heads and the author index.
%
\author{Jamal Al-Karaki$^{1,2,*}$
\and Muhammad Al-Zafar Khan$^{1,\dagger}$ \and Mostafa Mohamad$^{1,\ddagger}$ \and Dababrata Chowdhury$^{3,\diamondsuit}$}
\authorrunning{Security Challenges in Deep Learning Architectures}
% (feature abused for this document to repeat the title also on left hand pages)

% the affiliations are given next; don't give your e-mail address
% unless you accept that it will be published
\institute{$^{1}$College of Interdisciplinary Studies (CIS), Zayed University \\ Abu Dhabi, UAE\\
$^{2}$College of Engineering, The Hashemite University \\ Zarqa, Jordan \\
$^{3}$Christ Church Business School, Canterbury Christ Church University, \\ Kent, UK \\
$^{*}$\texttt{Jamal.Al-Karaki@zu.ac.ae} \\
$^{\dagger}$\texttt{Muhammad.Al-ZafarKhan@zu.ac.ae} \\
$^{\ddagger}$\texttt{Mostafa.Mohamad@zu.ac.ae} \\
$^{\diamondsuit}$\texttt{daba.chowdhury1@canterbury.ac.uk}
}

%
% NB: a more complex sample for affiliations and the mapping to the
% corresponding authors can be found in the file "llncs.dem"
% (search for the string "\mainmatter" where a contribution starts).
% "llncs.dem" accompanies the document class "llncs.cls".
%

\toctitle{Security Challenges in Deep Learning Architectures: Challenges and Countermeasures}
%\tocauthor{Authors' Instructions}
\maketitle

\begin{abstract}
 With the rise in the wholesale adoption of Deep Learning (DL) models in nearly all aspects of society, a unique set of challenges is imposed. Primarily centered around the architectures of these models, these risks pose a significant challenge, and addressing these challenges is key to their successful implementation and usage in the future. In this research, we present the security challenges associated with the current DL models deployed into production, as well as anticipate the challenges of future DL technologies based on the advancements in computing, AI, and hardware technologies. In addition, we propose risk mitigation techniques to inhibit these challenges and provide metrical evaluations to measure the effectiveness of these metrics. 
\end{abstract}

\section{Introduction}
\textit{Deep Learning} (DL), built upon ideas from mathematical optimization, is the art of mimicry of the human brain's ability to learn via information processing through iterative improvement in a multitude of layers. The more training examples the model has to work with -- that is, the more data the DL model ingests -- the more robust and predictable the model becomes. Due to the Universal Approximation Theorem \cite{hornik}, given a set of inputs, $\mathbf{x}=\left(x_{1},x_{2},\ldots,x_{n}\right)$ and outputs, $\mathbf{y}=\left(y_{1},y_{2},\ldots,y_{m}\right)$, for $m,n\in\mathbb{Z}^{+}$, with a sufficient amount of complexity (this arises due to having many neurons, connections, and nonlinear activations), DL models which have baked-in Artificial Neural Networks (ANNs) can learn to find the function $f:\mathbf{x}\to\mathbf{y}$, where $f$ is sufficiently a minimum; see \cite{lecun2015deep,goodfellow2016deep}. As a constantly evolving, disruptive, and pervasive technological tool, the rapid adoption of these models in fields like Natural Language Processing (NLP) -- most commonly associated with Large Language Models (LLMs) like ChatGPT \cite{achiam2023gpt} and Gemini \cite{team2023gemini} -- application domains in healthcare \cite{heimann2009statistical,wolfe2010american,radke2005image}, finance \cite{sezer2020financial,heaton2017deep}, and so on, and their incorporation into other Machine Learning (ML) subfields -- for example, in the field of Reinforcement Learning (RL), whereby Deep Neural Networks (DNNs)-powered algorithms learned the optimal state and action value functions to beat the world champion in Go \cite{silver2016mastering,silver2017mastering}, to protein folding \cite{senior2020improved}, to computational linear algebra \cite{fawzi2022discovering}. There are a plethora of security challenges that pose detrimental risks to the model developer, the organization, and the model adopters.  

While offering unprecedented capabilities and immense possibilities only limited by one's imagination, DL architectures are opaque, oftentimes called ``black boxes'' because human experts do not understand why certain decisions are taken by the model and are therefore susceptible to a range of security threats. Of particular prominence is the conscientious crafting of deceptive inputs in order to create incorrect outputs, and this, more so than others, underscores a critical susceptibility to security challenges. For example, in critical domains like autonomous vehicular control, if the model underwent a hostile attack, it could cause the vehicle to lose control and crash, seriously harming or even killing the occupants. 

As DL models become more widely adopted in commercial and industrial applications, the need for having a reliable understanding of the various risks and challenges is ever more important. In addition, understanding how to attenuate these risks is even more crucial. The challenges associated with DL can be directly directed to its architectures.

Further, with the widespread adoption of Cloud as a Service (CaaS) offerings, new challenges involve data privacy. Traditional methods like Federated Learning \cite{mcmahan2017communication}, Differential Privacy \cite{DworkMcSherry}, and Homomorphic Encryption \cite{GentryC} each have their own challenges. 

In this paper, we aim to shed light on the following research questions:
\begin{enumerate}
\item Is it possible to develop an all-embracing list of security challenges in the current DL architectures, and for future DL and AI architectures, can one provide a list of anticipated challenges?
\item Given a security challenge, can one provide a mitigation strategy against these risks?
\item Can one quantify the effectiveness of these strategies?
\end{enumerate}

While answering these questions may provide obvious, having such outlines in a clear, concise, and consolidated form does not exist, to the best of our knowledge. There are several case studies conducted by consulting firms that try to identify these risks and associated mitigation strategies, but these are confusing, more concentrated on AI holistically with less emphasis on DL, and contain a plethora of information that is unuseful. Thus, in this study, we aim to address this major hurdle.   

\subsection{Main Contributions}
In addressing the research questions, our foremost contributions of this research are as follows:
\begin{enumerate}
\item An exhaustive list of the current state-of-the-art challenges in DL security, together with discussions of how they work.
\item A new naming convention ($\alpha,\beta,\ldots$) that categorizes these attacks and places them into relevant buckets depending on their genus and type.
\item A comprehensive coverage of inhibitory strategies to mitigate the associated risk of each DL security challenge.  
\item The proposal of metrics on how to metricsize and quantify each of the attack vectors.
\item A demonstration of hypothetical scenarios on the mechanics of numerically using the metrics. 
\end{enumerate}

\noindent This paper is organized as follows:

In Sec. \ref{related works}, we discuss congruous research and their findings thereof.

In Sec. \ref{challenges}, we discuss the various challenges of DL model architectures, propose solutions to prevent them, and mitigate the associated risks. In addition, we provide metrics to ascertain the effectiveness of each of these proposed risk mitigation approaches. Lastly, we provide one sample calculation to demonstrate the mechanics of the metric. 

% In Sec. \ref{usage of metrics}, we provide hypothetical examples of fictitious data to demonstrate the mechanics of calculations with the metrics.

In Sec. \ref{conclusion}, we provide a recapitulation and reflection of our findings and propose future research directions.

\section{Related Work}\label{related works}

Below, we provide an abridged review of relevant literature pieces that helped influence our findings. This is not an exhaustive list.

In \cite{shokri2015privacy}, introduce a scheme of privacy-preserving collaborative training of DL models, a primitive form of Federated Learning. The key findings of this paper that relate to our research is that in order to successfully address the challenges associated with privacy during the training of DL models, one needs to consider: Selective parameter sharing instead of sharing each other's training data -- small subsets of the model weights and biases should be shared, and the exploitation of the parallelized nature of DL optimization schemes, such as the variants of gradient descent -- train models asynchronously and autonomously, and then conglomerate the model parameters which has the additional benefit of amplifying the efficiency of training and model scalability. 

In \cite{liu2020privacy}, the authors highlight various challenges in the form of a review piece associated with DL algorithms. In particular, the authors identified the following difficulties: Model theft and model reverse engineering, inference of sensitive training data (Federated Learning was specifically developed to combat this challenge), recovery of recognizable images in image-related learning tasks, and vulnerability to adversarial training examples, and highlight privacy-preserving techniques. The novelty of this research is the review of antipathetic attacks under physical conditions, which demonstrates practical utility, and relates the research to the real world. Similarly, in \cite{ha2020security}, the authors focus on the potential harm caused by false predictions and misclassifications of DL models and the graveness of protecting delicate information during the training and learning phases.  

In \cite{bae2018security}, the authors focus on how to ensure that DL models function optimally despite nefarious attacks, explore defenses against these attacks, and address threats associated with data privacy during model training. The novelty of these efforts is realized in the systematic characterization of security attack threats in the categories based on their timings: Poisoning attacks, which occur during training and corrupt the training data, and defense attacks, which occur during the inference phase of the model cycle, and impede the classification process. Analogously in \cite{papernot2016towards}, the authors systematize DL security and privacy and propose an overarching threat detection model that helps identify model vulnerabilities. The novelty associated with this piece of work is the formal exploration of the trade-off between model accuracy, model complexity, and sturdiness against adversarial exploit.

In \cite{rigaki2023garcia}, the author's study presents a taxonomy of privacy attacks in order to categorize different types of attacks based on adversarial knowledge and the assets being targeted. Further, the research identifies the causes of privacy leaks within ML models and systems and provides a review of the most common defenses proposed to mitigate these privacy attacks.

In \cite{LieXieGuoYangXiao}, the research focuses on DL safety associated with adversarial attacks in applications to 2D and 3D Computer Vision. By doing a comprehensive review of 170 research papers, they extend the idea of adversarial attacks beyond traditional imperceptible perturbations to be more diverse and complex than previously thought. 

In \cite{LeeYangVon}, the authors analyze 321 AI privacy incidents in order to provide a taxonomy of AI privacy risks. It was found that AI technologies, driven by DL models, can create new privacy risks, such as exposure risks from deepfake pornography, worsen existing privacy risks, such as surveillance risks due to the collection of training data, and can change the landscape of privacy concerns.

While we may include many other works, many of them are restatements of the research carried out in the aforementioned papers and simply discussing them here would be futile. We would also like to point out that our discussion of parallel research is by no means exhaustive. 

\section{Challenges and Proposed Solutions}\label{challenges}
In the subsections that follow, we discuss some of the major end-to-end challenges of DL architectures and how to mitigate them. 

\subsection{Attack Vectors and Challenges}
Using the methodical characterization provided in the literature, we have categorized the security challenges of DL architectures into the following groupings: Model Theft ($\alpha$), Antipathetic Attacks ($\beta$), Personal Data Privacy and Concealment ($\gamma$), Data Toxification ($\delta$), Explainability ($\epsilon$), Security Updates ($\zeta$), Deployment Difficulties ($\eta$), Novel Threats ($\theta$). Within these categories, we discuss the various types of security issues that could potentially arise. 

% In Fig. \ref{fig1}, we give a flowchart of the various risks according to our categorization, while in Fig. \ref{fig3}, we graphically render how extraction and inversion attacks occur.   

\begin{enumerate}
\item \textbf{Extraction Attacks:} $\alpha$. Intruders attempt to re- and deconstruct models by querying them and analyzing the outputs to steal their architecture or model parameters.  
\item \textbf{Larceny Attacks:} $\alpha$. The confidentiality of proprietary models is breached when intruders attempt the unauthorized wholesale replication of the model for iniquitous intent. 
\item \textbf{Inversion Attacks:} $\beta$. Sensitive information on training data is extricated by manipulating the model's predicted values.
\item \textbf{Elusory Attacks:} $\beta$. Input data is engineered to deceive the DL model, causing it to make erroneous predictions.
\item \textbf{Poisoning Attacks:} $\beta$. Introducing malicious training data in order to corrupt the model's training. 
\item \textbf{Data Dissipative Loss:} $\gamma$. Training data is involuntarily leaked during training and/or inference. 
\item \textbf{Membership Attacks:} $\gamma$. Intruders try various trial-and-error methods to determine whether a datapoint was part of the training data, thereby revealing other important information.
\item \textbf{Sub-rosa Attacks:} $\delta$. Model developers may have \textit{parti pris} -- ulterior motives and hidden agendas -- and embed hidden malicious actors in the DL model, having specific triggers in the training data to activate them. 
\item \textbf{DL Models are Blackboxes:} $\epsilon$. DL models are shrouded in opacity, making it difficult and oftentimes impossible to comprehend their decision-making processes and identify vulnerabilities. 
\item \textbf{Interpretability Dearth:} $\epsilon$. Even simple feed-forward network assemblies have a multitude of connectivity between nodes, thus making it extremely difficult, if not impossible, to understand what's going on and if there are any malicious actors deployed within them.
\item \textbf{Model Update Threats:}\label{model update threat} $\zeta$. These are vulnerabilities introduced during model updates. These can degrade the model's performance or outright compromise the model in its entirety. 
\item \textbf{Model Patching Threats:} $\zeta$. These are vulnerabilities introduced when small updates (patches) are introduced. Analogous to \ref{model update threat}, they can have comparable effects.
\item \textbf{Secure Deployment:} $\eta$. These are security risks, such as tampering or unwarranted access, that arise during the model deployment into the production phase.
\item \textbf{GenAI and Deepfakes:} $\theta$. GenAI and the associated underlying transformer-based models are constantly evolving. One of the major architectural threats is pretraining. If the previous model in which the current model is trained on has errors or interceptions, these could be inherited by the new model. Another major challenge is designing robust model architectures that can establish whether content is AI-generated and identify the source of deepfakes. This leads to the design of tamper-proof DL architectural systems for tracking content creation and modification. 
\item \textbf{Quantum AI Threats:} $\theta$. Quantum Computing, and Quantum Machine Learning, while they hold the potential to perform computations like never before by leveraging the Quantum Mechanical properties of Superposition and Entanglement, run the risk of receiving both classical and quantum attacks. One of the major challenges is developing architectures that are efficient in executing post-quantum cryptographic protocols for AI applications. Another major challenge is that quantum DL architectures can lose information along the way due to decoherence of the hardware with the environment.
\item \textbf{Knowledge Injection Attacks in Neurosymbolic AI Architectures:} $\theta$. These occur via the manipulating the symbolic knowledge base to influence system behavior. One of the major challenges is that for hybrid architectures, how does one verify the integrity of the symbolic knowledge? In addition, in order to enhance reasoning capabilities, how does one maintain privacy in order to preserve diverse knowledge.
\item \textbf{Challenges Associated with AGI Architectures:} $\theta$. We acknowledge that Artificial General Intelligence (AGI) is currently not a realized technology. However, we envisage that from a model architecture point of view, a major challenge would be to develop robust methods for specifying and enforcing ethical constraints in alignment with human values and intentions. In addition, due to the autonomous thought process of AGI, many experts have raised concerns about the self-modification of code and the associated security issues. These include constructing AGI model architectures that are self-tamper proof, effective architectural containment measures while not inhibiting self-determination, and developing architectural protocols for safe AGI-to-AGI interactions. 
\item \textbf{Challenges Associated with ASI Architectures:} $\theta$. The ability to supersede human intelligence with Artificial Superintelligence (ASI) seems even more far-fetched than the hopes of AGI. However, with the rapid AI revolution we are experiencing, it proves prudent to plan for such eventualities. One challenge is the development of ASI architectures that have built-in failsafe mechanisms that remain effective against superintelligent adversaries. Secondly, another major challenge is the creation of architectural security measures that cannot be trivially bypassed by ASI and the design of architectures that preserve human life from asymmetric intelligence scenarios.
\end{enumerate}

It is noteworthy to point out that for novel DL architectures, a common challenge is the preservation and security of information.

% \begin{figure}[t]
% \centering
% \includegraphics[width=1.2\linewidth]{images/a.pdf}
% \caption{Security risks encountered in DL model architectures}
% \label{fig1}
% \end{figure}

\subsection{Some Preventative Strategies and Risk Mitigation Policies}\label{preventative strategies}
In this subsection, we propose a set of inhibitory procedures and policies to reduce the risks associated with the security challenges posed by DL model architectures as follows:

% \begin{figure}[t]
% \centering
% \includegraphics[width=1.2\linewidth]{images/c.pdf}
% \caption{DL model extraction/inversion attacks}
% \label{fig3}
% \end{figure}

\begin{enumerate}
\item \textbf{Incorporation of Adversarial Training Examples:} In order to make the DL model robust, include adversarial examples in the training data so that the model is exposed to a plethora of datapoints that span both amicable and adversarial types. This increases the robustness of the overall model and makes it strong against attacks. Mathematically, we can quantify the adversarial robustness using
\begin{equation}
 R=\underset{x_{i}\in\mathbf{x}}{\arg\max}\;f(x_{i}^{\text{adv}}),   
\end{equation}
where $f(x_{i}^{\text{adv}})$ is the model prediction for adversarial example $x_{i}^{\text{adv}}$. The equation simply does a count of the number of correct classifications. 

% Suppose that in a dataset, there are 120 adversarial examples, and the model classifies 87 of them correctly. Then the model robustness is
% \begin{equation*}
% R=\frac{85}{120}\times100=70.83\%\;\text{(rounded off to two decimal places)}.
% \end{equation*}
% One may interpret this as the model being satisfactorily robust against adversarial examples. 

\item \textbf{Noise Filtering of Raw Data:} During model preprocessing, apply noise filters to sieve out adversarial actors that may be hidden within the data. Mathematically, we borrow a tool from signal processing; we can quantify the quality of the filtered data using
\begin{equation}
DQ_{\text{filtered}}=1-\frac{\mathcal{N}_{\text{after}}}{\mathcal{N}_{\text{before}}}    
\end{equation}
where $\mathcal{N}_{\text{before}}$ is the number of datapoints before filtering, and $\mathcal{N}_{\text{after}}$ is the number of datapoints that remains after filtering.

\item \textbf{Data Autoclaving:} Sanitize your raw data by cleaning it in order to remove anomalous examples. We quantify the effectiveness of cleaning via the anomaly detection rate
\begin{equation}
A=\frac{n_{A}}{m},
\end{equation}
where $n_{A}$ is the total number of anomalies detected, and $m$ is the total number of datapoints.

\item \textbf{Model Access Control:} Have restricted model access control policies and security clearance protocols in place whereby only certain user classes can have access to the model and the training data. Mathematically, the effectiveness of the model access control can be calculated using
\begin{equation}
A/C=\frac{N_{\text{unauthorized}}}{N},    
\end{equation}
where $N_{\text{unauthorized}}$ is the number of unauthorized access attempts that were blocked, and $N$ is the total number of access attempts.

\item \textbf{Protection from using Output to Surmise Input:} Use privacy-preserving techniques, such as differential privacy, so that the model output does not reveal information about the model's input. In order to mathematically quantify privacy preservation, we set a privacy budget threshold, $\mathcal{E}$, and calculate the total privacy losses. If the total privacy loss is less than the threshold, then the privacy-preserving technique was effective. Mathematically, we require the privacy loss, $P$, to be at most
\begin{equation}
P\leq\mathcal{E}.   
\end{equation}

\item \textbf{Model Surveillance:} Monitor model queries and the number of queries per user to detect anomalous model queries. Analogous to our proposal for quantifying the effectiveness of data autoclaving, we measure the query anomaly detection rate as
\begin{equation}
Q=\frac{q_{A}}{q},
\end{equation}
where $q_{A}$ is the total number of anomalous queries detected, and $q$ is the total number of queries.
\end{enumerate}

\begin{table}[H]
\centering
\begin{tabular}{|p{4cm}|p{11cm}|}
\hline
\textbf{Metric} &\textbf{Calculation} \\
\hline 
 Adversarial Training &Suppose that in a dataset, there are 120 adversarial examples, and the model classifies 87 of them correctly. Then the model robustness is
\begin{equation*}
R=\frac{85}{120}\times100=70.83\%\;\text{(rounded off to two decimal places)}.
\end{equation*}
One may interpret this as the model being satisfactorily robust against adversarial examples.  \\
\hline 
Noise Filtering &Suppose that a dataset contains $2\;000$ examples, and after applying a noise filter $\mathcal{F}$ that removes all those datapoints where $x_{i}\geq\frac{7}{2}\bar{x}$, i.e. 3.5 times greater than the mean, it was determined that 87 noisy datapoints still remained. Then, the quality of the filtered data is
\begin{equation*}
DQ_{\text{filtered}}=\left(1-\frac{80}{2\;000}\right)\times100=95.65\%\;\text{(rounded off to two decimal places)}.
\end{equation*}
One may interpret this as the filter did an effective job of removing anomalies. However, one can apply more filters to refine the data even further or adjust the filtering criteria to be more stringent or relaxed based on domain knowledge. \\   
\hline
Data Autoclaving &Suppose that a dataset has $10\;000$ examples, and 35 anomalies were detected. Then, the anomaly detection rate is
\begin{equation*}
A=\frac{35}{10\;000}\times100=0.35\%\;\text{(rounded off to two decimal places)}.    
\end{equation*}
The rate is poorly low, which one may interpret as needed to re-examine the dataset to search for more anomalies. \\
\hline 
\end{tabular}
\caption{Demonstration of the mechanics of some metrics}
\label{tab100}
\end{table}

In Tab. \ref{tab100}, we provide some sample calculations to demonstrate how the proposed metrics work on scenarios that are reminiscent of reality.

\section{Conclusion}\label{conclusion}
We have considered the various security challenges that DL models, and thereby their architectures, can possess, together with strategies and policies to mitigate these risks and metrical measures that can quantify the effectiveness of these strategies. This work was built upon characterizations provided in the literature, and we meticulously filtered them down to address individual security risks posed. It is our belief that our characterization is unique, and can greatly assist in the planning and strategizing by policymakers. As a future consideration, it would prove a worthwhile task to collect data, or generate synthetic data, and apply ML and DL models to predict the various risk types -- that is, ``using DL to detect DL security threats''; this is a venture that will be considered in future works. In doing so, this will help strategists and risk managers to design effective master plans to overcome this \textit{faux pas}.

%===============================================================

\end{document}